\documentclass[conference]{IEEEtran}
\IEEEoverridecommandlockouts
\usepackage{cite}
\usepackage{url}
\usepackage{amsmath,amssymb,amsfonts}
\usepackage{algorithmic}
\usepackage{graphicx}
\usepackage{textcomp}
\usepackage{xcolor}
\def\BibTeX{{\rm B\kern-.05em{\sc i\kern-.025em b}\kern-.08em
    T\kern-.1667em\lower.7ex\hbox{E}\kern-.125emX}}

\usepackage[ruled,vlined]{algorithm2e}
\usepackage{booktabs}

\usepackage{amsthm}

\begin{document}

\title{Over the Air Computing for \\Satellite Networks in 6G\\
\thanks{This work has received funding by the Spanish ministry of science and innovation under project IRENE (PID2020-115323RB-C31) funded by MCIN/AEI/10.13039/501100011033.}
}

\author{\IEEEauthorblockN{Marc Martinez-Gost\IEEEauthorrefmark{1}\IEEEauthorrefmark{2}, Ana Pérez-Neira\IEEEauthorrefmark{1}\IEEEauthorrefmark{2}}
\IEEEauthorblockA{
\IEEEauthorrefmark{1}Centre Tecnològic de Telecomunicacions de Catalunya, Spain\\
\IEEEauthorrefmark{2}Dept. of Signal Theory and Communications, Universitat Politècnica de Catalunya, Spain\\
Emails: \{mmartinez, aperez\}@cttc.es}
}

\maketitle

\begin{abstract}
6G and beyond networks will merge communication and computation capabilities in order to adapt to changes. As they will consist of many sensors gathering information from its environment, new schemes for managing these large amounts of data are needed. For this purpose, we review Over the Air (OTA) computing in the context of estimation and detection. For distributed scenarios, such as a Wireless Sensor Network, it has been proven that a separation theorem does not necessarily hold, whereas analog schemes may outperform digital designs. We outline existing gaps in the literature, evincing that current state of the art requires a theoretical framework based on analog and hybrid digital-analog schemes that will boost the evolution of OTA computing.
Furthermore, we motivate the development of 3D networks based on OTA schemes, where satellites function as sensors. We discuss its integration within the satellite segment, delineate current challenges and present a variety of use cases that benefit from OTA computing in 3D networks.
\end{abstract}


\section{Introduction}

Next-generation networks will join communication and computing since a large number of sensors will be deployed to enable the development of intelligent networks capable of responding to the environment and take appropriate decisions. These sensors take measurements over which compute calculations. In terms of efficiency, these computations will be carried in a distributed fashion, complementing the communication side of the network.

It is well known that for architectures beyond the point-to-point setting, discerning between source and channel coding may be suboptimal. Furthermore, there has been an increasing interest in the development of analog schemes, as envisaged by \cite{analog}.
New physical layer designs need to be developed to manage the massive volumes of sensed data. Besides, these have to consider the objectives of the computing tasks, and not only the communication goals. In other words, there is a shift from considering only rate-oriented performance indicators to contemplating task-oriented targets (e.g., minimize a cost that depends on the data). In this sense, Over the Air (OTA) computing is a recently new paradigm that takes advantage of the additive nature of the Multiple Access Channel (MAC). In an OTA scheme, all transmitters send signals coherently such that the receiver obtains a unique overlapped signal. In Wireless Sensor Networks (WSN) this observation is convenient because the receiver (e.g., fusion center) is not particularly interested in acquiring all the readings, but a function of them. We review the state of the art in OTA for estimation and detection problems, revealing that most of the literature proposes solutions based on ad-hoc designs because no general tools have been envisioned to design analog architectures.

The development of OTA schemes for WSN can be envisaged in a 3D network, where satellites function as sensors gathering data and they transmit it concurrently towards a ground fusion center. Nevertheless, apart from the challenges inherited from OTA, its deployment over the satellite segment exhibits new challenges. Because of the density of satellites, interferences can happen within satellites from the same constellation, within the same orbit or even different orbits. Contrarily to research focused on mitigating interferences \cite{dense_sky}, depending on the service and application, these interferences can be used for the reception benefit with the OTA strategy.

In this paper we review OTA computing over WSN in the context of estimation and detection scenarios, while indicating the gaps that exist in the literature, for which we provide future lines of research. Additionally, we show the feasibility of deploying OTA schemes over 3D networks, including numerous use cases that can take advantage of it.

The remaining part of the paper proceeds as follows: section II begins with an overview of OTA computing and examines its integration in distributed estimation and detection problems; section III discusses the development of OTA schemes in 3D networks and outlines current challenges; section IV presents use cases and section V summarizes the paper and proposes future lines of research.

\section{Over the Air Computing}
In a point-to-point setting Shannon proved that it is optimal to decompose the process of communication in two disjoint phases: once the source is sampled, a source code is used to compress the information; then, a channel code is applied for it to be resilient to noise \cite{shannon}. These stages are independent in the sense that the latter has no information regarding the source, the only link is due to rate constraints. This is known as the \textit{source-channel separation theorem}, and pushed the development of communication systems in the digital domain. 

\begin{figure}[t]
\centering
\includegraphics[width=\columnwidth]{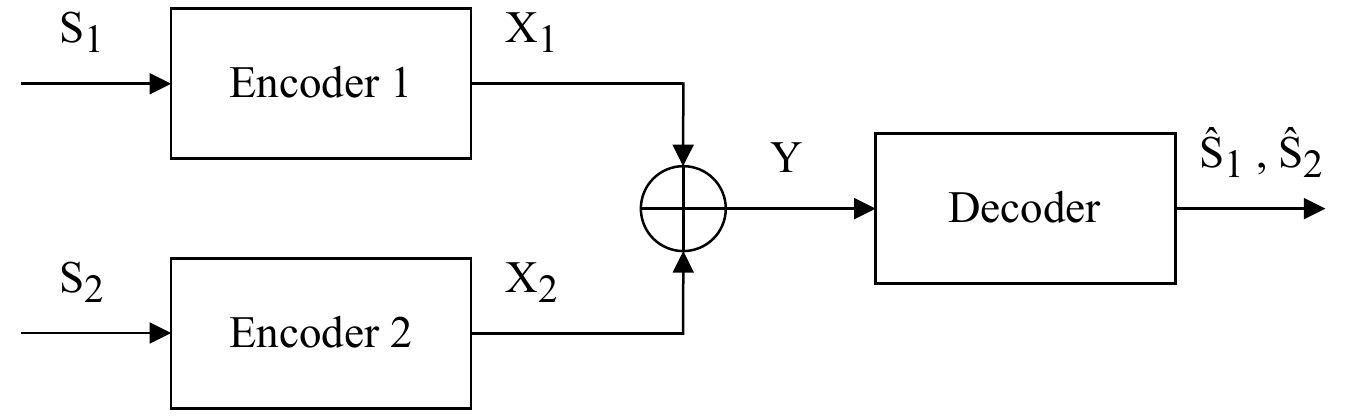}
\caption{2-user Adder MAC.}
\label{fig: 2_mac}
\end{figure}

Nonetheless, a WSN differs from the previous setup for two main reasons: The first one is that a WSN is a multipoint-to-point scenario. In \cite{analog} the following 2-user Adder MAC is proposed to show how a simple analog scheme can outperform a standard digital one. Consider two independent sources $S_1$ and $S_2$, with probabilities $P(S_1=0,S_2=0)=P(S_1=0,S_2=1)=P(S_1=1,S_2=1)$ and $P(S_1=1,S_2=0)=0$, depicted in Figure \ref{fig: 2_mac}. The rate-distortion and capacity regions do not intersect, evincing that there is no digital code providing rates for which the receiver can recover both sources with no distortion. Conversely, as the channel matches the structure of the sources, the analog approach of transmitting the uncoded signals allows the receiver to determine the original sources unequivocally. This example goes to show that a separation theorem does not hold in general for WSN, whereas a joint source-channel code is needed for optimal performance.

The second reason stems from the small-sized nature of the sensed data. For instance, consider a WSN recording temperature measurements. In an distributed estimation problem, each node senses a temperature readout and the objective is the receiver (e.g., fusion center) to obtain an estimation from all the distributed data. Contrarily, in a distributed detection problem, each sensor gathers a measurement and runs a local hypothesis test (e.g., to check whether it is above a threshold). Then, the receiver gathers the local decisions and computes a global decision. In \cite{ota}, the authors define the computation rate as the ratio between the input and output number of bits that a computation code produces, $\kappa=k/n$. When $\kappa=1$, it breaks with the traditional framework proposed by Shannon, because asymptotically infinite-length codes are not allowed. As seen in both estimation and detection problems, the size of the data is small enough to consider the transmission of small packets and, thus, the asymptotic conditions under which the channel capacity is defined do no longer hold.

These considerations motivate the development of new schemes for WSN. Namely, Over-the-Air computing is a new paradigm to reliable compute functions over the MAC channel \cite{ota}. Considering that sensors gather vast amounts of data and that the receiver is not interested in acquiring all instances, rather a function of them (e.g., mean), the additive property of the MAC channel is exploited for the receiver to harness interferences and obtain a unique signal. Notice that this also reduces the communication needs of the system because no resources are allocated to achieve orthogonalization.

The theory of OTA systems resides on the theorem proved by Kolmogorov \cite{kolmogorov}, stating that any continuous multivariate function can be expressed as a superposition of only one variable functions,
\begin{align}
    f(s_1,\dots, s_N) = g\left( \sum_{n=1}^{N} h_n(s_n)  \right),
    \label{eq: kolmogorov}
\end{align}
where $f(\cdot)$ is called a nomographic function. The set of functions $\{h_n\}_{n=1}^{N}$ are called pre-processing functions and are computed in a distributed fashion at each transmitter, while $g(\cdot)$ is a post-processing function computed at the receiver over the superposition of transmitted signals.

The simultaneous transmission provided by OTA computing can be advantageous, for instance, to reduce the complexity of the receiver since it does not scale with the number of sensors, this is, the receiver acquires a unique signal at its input. Because of this, the receiver cannot compensate each one of the channels from every sensor. In this respect, OTA computing involves different signal processing techniques with respect to a Parallel Access Channel (PAC) constructed with orthogonal resources. For instance, a multi-antenna receiver cannot perform zero-forcing channel equalization, at most it can generate a multicast beamformer that aims to point towards all the sensors. This is the design presented in \cite{bengttson}, where the beamforming vector is constructed so that all effective channels have similar amplitude. On the other hand, the channel compensation has to be performed at each transmitter, as it is the only location where the signals are decoupled. This was also analysed in \cite{bengttson}, where the transmit power allocation at the sensors is designed in order to minimize the power consumption  while  constraining  the  SINR  to  be  above  a  threshold. In \cite{scaling_laws}, the  power allocation at the sensors and at the receiver are designed to minimize the mean squared error (MSE) between the local readings and the superimposed signal at the receiver.

OTA computing manages different diversity schemes, because not only channels are different, but also each signal may be different from the rest. This happens because the observations are subject to noise and the spatial distribution of the nodes may imply obtaining different readouts. Notice this is more complex than a Frequency Diversity Spread Spectrum (FDSS) scheme because, not only each sample is sent over a different non-overlapping frequency band, but it is assumed that all transmitters have the same symbol \cite{FDSS}.

\subsection{OTA for distributed estimation}
In distributed estimation, nodes gather measurements over which a parameter is estimated. In general, the Maximum Likelihood (ML) estimator is chosen. Notice that the objective in estimation is to minimize the MSE between the readings and the estimated parameter. When it is not possible to compute it in a closed form, an iterative process is implemented, where interactions (i.e., communication and computations) among them is performed until a global estimation is reached. Notice that these interactions can be performed among nodes or with respect to the fusion center. In the latter it is recognizable that OTA computing can ease the convergence of the algorithm. This is the main goal of Federated Learning algorithms, which try to come up with a global model (e.g., neural network) by means of distributed local models. We refer the reader to \cite{FL} for an extensive up-to-date review of learning algorithms over wireless networks.

\begin{figure}[t]
\centering
\includegraphics[width=\columnwidth]{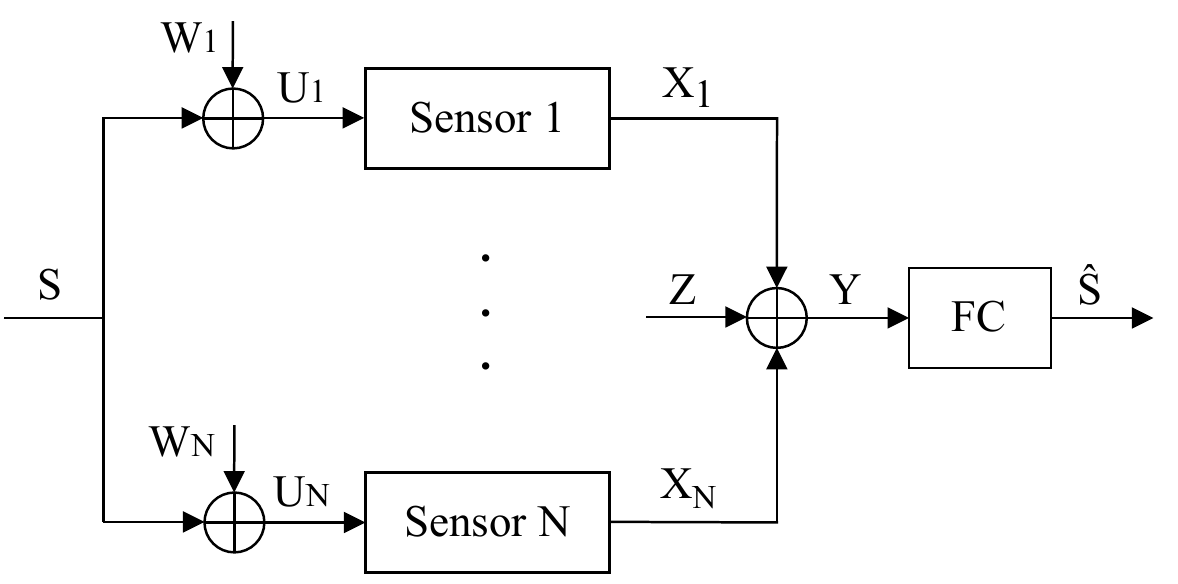}
\caption{WSN of $N$ sensors and a Fusion Center (FC) for parameter estimation.}
\label{fig: refining sensor network}
\end{figure}

The authors in \cite{analog} analyse the sensor network architecture illustrated in Figure \ref{fig: refining sensor network}, where sources follow a Gaussian distribution and for which an analog scheme performs exponentially better than the digital scheme in a scaling sense. For a digital encoding, it corresponds to the quadratic Gaussian Chief Executive Officer (CEO) problem \cite{CEO}. In this case, the distortion $D$ as a function of the number of sensors scales as in (\ref{eq: distortion_digital}), where $P_{tot}$ corresponds to the total available power in transmission for $N$ sensors.
\begin{equation}
    D_{digital}(N)\sim \frac{1}{log\left(NP_{tot}\right)}
    \label{eq: distortion_digital}
\end{equation}

In the analog architecture, the signal is scaled by an appropriate power factor that depends on the number of sensors, the transmission power, and the power level of the signal and noise. Since the received signal remains Gaussian, the optimal estimator of $S$ corresponds to $\hat{S}=\left(E\{SY\}/E\{Y^2\}\right)Y$ \cite{analog}. The distortion scales as in (\ref{eq: distortion_analog}), revealing that the digital architecture requires exponentially more sensors to attain the same distortion level. Therefore, for distributed estimation with poor communication infrastructure, there is no separation theorem.

\begin{equation}
    D_{analog}(N)\sim \frac{1}{NP_{tot}}
    \label{eq: distortion_analog}
\end{equation}

For these general networks, no separation theorem holds, which demands the development of new designs for either analog or hybrid digital-analog schemes. We will refer to them as Radio Frequency Coding (RFC) techniques. RFC does not operate with Galois fields and aim to provide new waveform designs that adapt to the needs of joint computing and communication \cite{RFC}.

With respect to estimation, the Cramér-Rao Bound (CRB) works as a lower bound for the variance of an estimator. It has been shown that the quantization process of the measurements affects the parameter estimate, which adds another layer of complexity. In \cite{CR_bound}, the CRB is derived for a general quantizer and signal model, which can be a path to follow for a suitable waveform coding design.

For saving communication bandwidth, computations can be performed locally at the sensors. When, instead of working with parameters, we work with associated hypothesis, an estimation problem turns into a detection problem.

\subsection{OTA for distributed detection}
In a detection problem there is a set of hypothesis $\{\mathcal{H}_i;\ i=1,\dots,K\}$ that indicate the presence or absence of one or several events. In a decentralized setting, each sensors gathers data over which runs a decision criterion. Then all nodes transmit their decisions coherently towards the receiver, in which a global decision is computed. In detection problems the objective is reducing the probability error, associated to false alarm or miss events.

The transmission of hypothesis in contrast to raw data is advantageous for several reasons: 
1) It is not straightforward how to adapt the OTA scheme to certain types of sources. For example, in the case of images, the fusion center would receive a superimposed image, which may not even be useful if the pictures were taken from different perspectives and distances. Conversely, the hypothesis are well defined and do not have any dependency on the medium as the data may have; 2) the fusion center is offloaded by computing hypothesis locally, and 3) because of its binary nature, hypothesis reduce the size of the data and, thus, reduces the communication needs of the system.

\begin{figure}[t]
\centering
\includegraphics[width=\columnwidth]{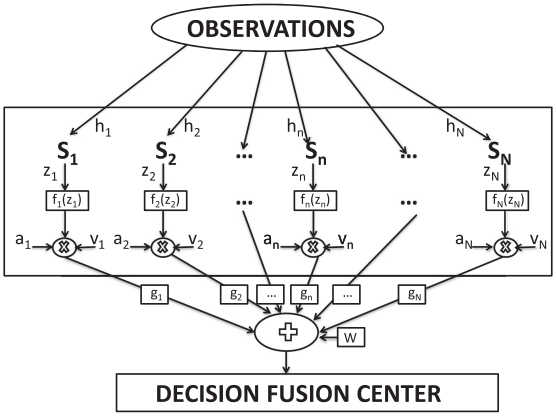}
\caption{OTA scheme for detection. (Source: \cite{bengttson}).}
\label{fig: ota_scheme}
\end{figure}

An OTA scheme for detection is depicted in Figure \ref{fig: ota_scheme}, where sensor $n$ gathers a noisy version $h_n$ of the source $S$ and computes a local decision from a hypothesis test, $z_n=\theta_n(h_n)$. Then, these decisions are digitally modulated with a function $f_n(\cdot)$, amplified with $a_n$ and phase shifted with $v_n$. The noise at the receiver is $W$ and $g_n$ is channel from the $n$-th sensor to the receiver. 

The nature of hypothesis in detection problems involves working with finite codebooks, which differs from estimation problems. As an example, in \cite{bengttson} the authors propose an ad-hoc approach, where PSK constellations are used depending on the number of hypothesis that a sensor can distinguish (e.g., a BPSK scheme for two hypothesis, whereas four hypothesis are mapped to a QPSK). On the other hand, lattice codes have gained relevance for OTA schemes because they can achieve capacity on an AWGN channel \cite{ota}. It is still to be explored if there are analog or hybrid digital-analog schemes that outperform this kind of digital schemes.

\section{Integration of OTA in 3D networks}
The development of 3D networks falls under the definition of the OTA paradigm, when satellites operate as sensors gathering data from the Earth and they send it concurrently towards the on-ground receiver. Additionally, the use of small data packets in WSN matches the latency requirements of satellite to ground communications. In this distributed scenario, data may need to be preprocessed before transmission in order to save communication bandwidth, which can be performed via the on-board computation capabilities of the satellites. Notice that this upgrades the operational means of the network beyond a communication system, incorporating computational capabilities as well. Similarly, the uplink case can be considered as well, where IoT devices communicate simultaneously with a satellite.

The consideration of 3D networks as a WSN comprises several challenges due to its heterogeneity between the ground and satellite segments. Besides, the inclusion of OTA computing brings a set of challenges as well that need to be addressed towards the development of 3D networks in 6G:
\begin{itemize}
    \item \textit{Synchronization}: because simultaneous transmission is needed for OTA computing and satellites are located at different distances, synchronization has to be assessed. AirShare is a system in which a clock is transmitted and shared for all sensors for coherent transmissions \cite{airshare}. In another line, \cite{coarse_sync} develops an OTA analog scheme that supports coarse-block synchronization. Synchronization is even more demanding for a dynamic topology as a satellite constellation.
    \item \textit{Channel State Information}: it is not feasible for each sensor to acquire knowledge of the channel. WSN would profoundly benefit from the development of blind schemes. Several alternatives have been developed to avoid massive CSI for OTA schemes \cite{CSI_1, CSI_2}.
\end{itemize}

\subsection{Constellation Design}
For sensing services, the density of satellites must be large enough to provide global coverage. LEO satellites are more suitable because their closeness to the Earth reduces the propagation delay and attenuation. The latter is crucial for IoT-oriented applications, since ground devices have stringent energy consumption requirements. Besides that, for Earth Observation services, LEO satellites can obtain better resolution with respect to MEO or GEO satellites.

\begin{figure}[t]
\centering
\includegraphics[width=\columnwidth]{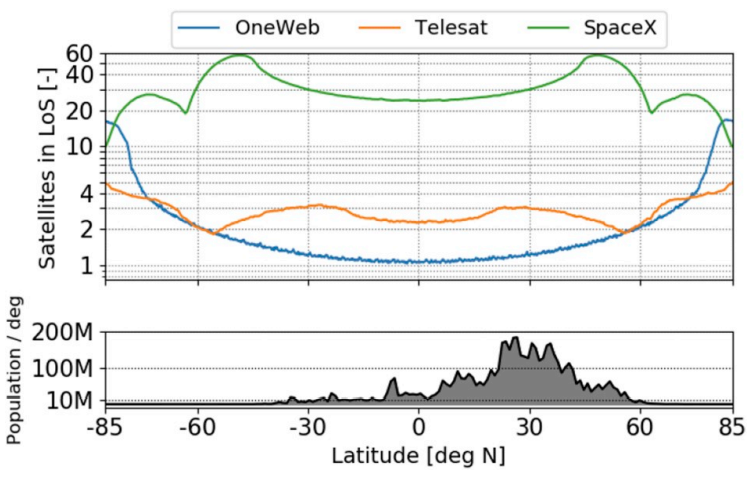}
\caption{Number of satellites in Line of Sight vs. latitude. (Source: \cite{portillo}).}
\label{fig: coverage_los}
\end{figure}

Because of the different needs, constellations designed for broadband communication may not comply with the requirements of sensing applications. For instance, Starlink's constellation is designed to have more satellites covering the most populated regions, for which inclined orbits are used \cite{portillo}. However, these orbits do not cover the polar regions, which may be of interest for sensing applications.
Besides, in order to take advantage of OTA computing (downlink case), several satellites need to cover the same region and, specifically, point towards the same receiver. This can be accomplished by means of  neighboring satellites within the same constellation and ring, satellites from the same constellation and different shells and satellites from different orbits. These scenarios increase in complexity as the network is more heterogeneous (e.g., synchronization issues). In \cite{portillo} the authors study how many satellites are in Line-of-Sight (LoS) depending on the latitude and for different constellations. As Figure \ref{fig: coverage_los} shows, the Starlink constellation provides the densest number of satellites, whereas Oneweb may not be enough. This raises the questions whether new constellation designs have to be developed for distributed sensing integrating OTA communications or these broadband constellations could be reused for different purposes.

\section{Use Cases}
In order to exploit the satellite segment for distributed sensing, each satellite must obtain data and use the aggregation property of the channel medium to ease computation. As previously discussed, the computation capabilities of satellites can also be exploited by performing on-board computations. In sensing applications the receiver is interested in the latest possible update of the information. This reinforces the idea of Age of Information, in which the significance shifts towards the timeliness of the information \cite{age_info}. We distinguish between the following categories:
\begin{itemize}
    \item \textbf{Downlink}: real-time tracking of containers in logistics and real-time maritime surveillance \cite{paz} are two use cases that can be enhanced by the use of satellites, since it is not feasible to deploy a ground infrastructure. Besides, the large amount of satellites allows to update the status in a timely manner.\\
    Another use case is environmental monitoring, such as controlling coal emissions, and assessing the dose and coverage of pesticides \cite{pesticides}.
    
    \item \textbf{Uplink}: extending the use of the AIS system \cite{AIS}, several ships can communicate with a satellite, where the latter can be used a relay node.
\end{itemize}

\section{Conclusions}
In this paper we have presented the ideas that motivate the development of OTA computing for 3D networks in 6G. With respect to OTA schemes, we have analysed two general problems, namely, estimation and detection. We have shown the benefits of analog-based schemes and indicate that the current state of the art lacks a theoretical frameworks that unifies the coding and waveform designs for OTA computing. As regards to its inclusion in 3D networks, we have exhibited its feasibility, outlined up-to-date challenges and proposed a variety of use cases that benefit from OTA strategies.

\section{Acknowledgement}
The authors would like to acknowledge helpful discussions with Israel Leiva-Mayorga that assisted the research.

This work has been performed under a programme funded by the ICREA Academia.

\bibliographystyle{ieeetr}
\bibliography{refs}

\begin{thebibliography}{10}

\bibitem{analog}
M.~Gastpar, M.~Vetterli, and P.~Dragotti, ``Sensing reality and communicating
  bits: a dangerous liaison,'' {\em IEEE Signal Processing Magazine}, vol.~23,
  no.~4, pp.~70--83, 2006.

\bibitem{dense_sky}
J.~Suilmann, A.~M. Voicu, L.~Simić, and P.~Mähönen, ``The dense sky:
  Evaluating system coexistence of new ngso satellite constellations in the ka
  band,'' in {\em 2021 IEEE Globecom Workshops (GC Wkshps)}, pp.~1--6, 2021.

\bibitem{shannon}
C.~E. Shannon, ``A mathematical theory of communication,'' {\em The Bell System
  Technical Journal}, vol.~27, no.~3, pp.~379--423, 1948.

\bibitem{ota}
B.~Nazer and M.~Gastpar, ``Computation over multiple-access channels,'' {\em
  IEEE Transactions on Information Theory}, vol.~53, no.~10, pp.~3498--3516,
  2007.

\bibitem{kolmogorov}
A.~N. Kolmogorov, ``On the representation of continuous functions of many
  variables by superposition of continuous functions of one variable and
  addition,'' {\em Dokl. Akad. Nauk SSSR}, vol.~114, no.~5, pp.~953--956, 1957.

\bibitem{bengttson}
S.~R. Panigrahi, N.~Björsell, and M.~Bengtsson, ``Data fusion in the air with
  non-identical wireless sensors,'' {\em IEEE Transactions on Signal and
  Information Processing over Networks}, vol.~5, no.~4, pp.~646--656, 2019.

\bibitem{scaling_laws}
W.~Liu, X.~Zang, Y.~Li, and B.~Vucetic, ``Over-the-air computation systems:
  Optimization, analysis and scaling laws,'' {\em IEEE Transactions on Wireless
  Communications}, vol.~19, no.~8, pp.~5488--5502, 2020.

\bibitem{FDSS}
G.~Kaleh, ``Frequency-diversity spread-spectrum communication system to counter
  bandlimited gaussian interference,'' {\em IEEE Transactions on
  Communications}, vol.~44, no.~7, pp.~886--893, 1996.

\bibitem{FL}
M.~Chen, D.~Gündüz, K.~Huang, W.~Saad, M.~Bennis, A.~V. Feljan, and H.~V.
  Poor, ``Distributed learning in wireless networks: Recent progress and future
  challenges,'' 2021.

\bibitem{CEO}
T.~Berger, Z.~Zhang, and H.~Viswanathan, ``The ceo problem [multiterminal
  source coding],'' {\em IEEE Transactions on Information Theory}, vol.~42,
  no.~3, pp.~887--902, 1996.

\bibitem{RFC}
A.~Perez-Neira, ``Radio frequency coding,'' in {\em 2019 International
  Conference on Electromagnetics in Advanced Applications (ICEAA)},
  pp.~0665--0666, 2019.

\bibitem{CR_bound}
P.~Stoica, X.~Shang, and Y.~Cheng, ``The cramér–rao bound for signal
  parameter estimation from quantized data [lecture notes],'' {\em IEEE Signal
  Processing Magazine}, vol.~39, no.~1, pp.~118--125, 2022.

\bibitem{airshare}
O.~Abari, H.~Rahul, D.~Katabi, and M.~Pant, ``Airshare: Distributed coherent
  transmission made seamless,'' in {\em 2015 IEEE Conference on Computer
  Communications (INFOCOM)}, pp.~1742--1750, 2015.

\bibitem{coarse_sync}
M.~Goldenbaum and S.~Stanczak, ``Robust analog function computation via
  wireless multiple-access channels,'' {\em IEEE Transactions on
  Communications}, vol.~61, no.~9, pp.~3863--3877, 2013.

\bibitem{CSI_1}
M.~M. Amiri, T.~M. Duman, and D.~Gündüz, ``Collaborative machine learning at
  the wireless edge with blind transmitters,'' in {\em 2019 IEEE Global
  Conference on Signal and Information Processing (GlobalSIP)}, pp.~1--5, 2019.

\bibitem{CSI_2}
L.~Chen, N.~Zhao, Y.~Chen, F.~R. Yu, and G.~Wei, ``Over-the-air computation for
  iot networks: Computing multiple functions with antenna arrays,'' {\em IEEE
  Internet of Things Journal}, vol.~5, no.~6, pp.~5296--5306, 2018.

\bibitem{portillo}
I.~{del Portillo}, B.~G. Cameron, and E.~F. Crawley, ``A technical comparison
  of three low earth orbit satellite constellation systems to provide global
  broadband,'' {\em Acta Astronautica}, vol.~159, pp.~123--135, 2019.

\bibitem{age_info}
B.~Soret, I.~Leyva-Mayorga, S.~Cioni, and P.~Popovski, ``5g satellite networks
  for internet of things: Offloading and backhauling,'' {\em International
  Journal of Satellite Communications and Networking}, vol.~39, no.~4,
  pp.~431--444, 2021.

\bibitem{paz}
{European Space Agency}, ``{PAZ SAR satellite mission of Spain}.''
  \url{earth.esa.int/web/eoportal/satellite-missions/p/paz}, 2021.

\bibitem{pesticides}
P.~Flémal, O.~Pigeon, P.~Dardenne, J.~F. Pierna, V.~Baeten, and P.~Vermeulen,
  ``Assessment of pesticide coating on cereal seeds by near infrared
  hyperspectral imaging,'' {\em Journal of Spectral Imaging}, vol.~6, no.~1,
  p.~a1, 2017.

\bibitem{AIS}
A.~Harati-Mokhtari, A.~Wall, P.~Brooks, and J.~Wang, ``Automatic identification
  system (ais): data reliability and human error implications,'' {\em The
  Journal of Navigation}, vol.~60, no.~3, p.~373, 2007.

\end{thebibliography}

\end{document}